\documentclass[sigconf, screen, authorversion]{acmart}
\usepackage{multirow}
\usepackage{pifont}
\usepackage{enumitem}
\usepackage{dirtytalk}

\newcommand{\Circle}{\ensuremath{\raisebox{-0.5ex}{\scalebox{2.0}{$\circ$}}}}
\newcommand{\CIRCLE}{\ensuremath{\raisebox{-0.5ex}{\scalebox{2.0}{$\bullet$}}}}

\AtBeginDocument{%
  }

\setcopyright{acmlicensed}
\copyrightyear{2025}
\acmYear{2025}
\acmDOI{XXXXXXX.XXXXXXX}

\acmConference[UKICER ’25]{2025 Conference on United Kingdom \& Ireland Computing Education Research}{September 04--05,
  2025}{Edinburgh, UK}

\acmISBN{978-1-4503-XXXX-X/18/06}

\begin{document}

\title[Student Engagement with GenAI Feedback]{That's Not the Feedback I Need! - Student Engagement with GenAI Feedback in the Tutor Kai}

\author{Sven Jacobs}
\orcid{0009-0000-5079-7941}
\affiliation{%
  \institution{University of Siegen}
  \city{Siegen}
  \country{Germany}
}
\email{sven.jacobs@uni-siegen.de}

\author{Maurice Kempf}
\orcid{0009-0000-8860-2353}
\affiliation{
  \institution{University of Siegen}
  \city{Siegen}
  \country{Germany}
}
\email{maurice.kempf@uni-siegen.de}

\author{Natalie Kiesler}
\orcid{0000-0002-6843-2729}
\affiliation{
   \institution{Nuremberg Tech}
   \city{Nuremberg}
   \country{Germany}
}
\email{natalie.kiesler@th-nuernberg.de}

\renewcommand{\shortauthors}{Jacobs, Kempf, Kiesler}

\begin{abstract}
The potential of Generative AI (GenAI) for generating feedback in computing education has been the subject of numerous studies. However, there is still limited research on how computing students engage with this feedback and to what extent it supports their problem-solving. 
For this reason, we built a custom web application providing students with Python programming tasks, a code editor, GenAI feedback, and compiler feedback.  
Via a think-aloud protocol including eye-tracking and a post-interview with 11 undergraduate students, we investigate (1) how much attention the generated feedback received from learners and (2) to what extent the generated feedback is helpful (or not).
In addition, students' attention to GenAI feedback is compared with that towards the compiler feedback. We further investigate differences between students with and without prior programming experience. 
The findings indicate that GenAI feedback generally receives a lot of visual attention, with inexperienced students spending twice as much fixation time. 
More experienced students requested GenAI less frequently, and could utilize it better to solve the given problem. It was more challenging for inexperienced students to do so, as they could not always comprehend the GenAI feedback. They often relied solely on the GenAI feedback, while compiler feedback was not read. 
Understanding students' attention and perception toward GenAI feedback is crucial for developing educational tools that support student learning.
\end{abstract}




\maketitle

\section{Introduction}
\label{sec:introduction}
Learning to program is cognitively complex, and students require support throughout their studies, e.g., via individualized and timely feedback~\cite{jeuring.2022}. However, this task is becoming increasingly challenging for educators, especially in large university courses with a diverse student body. Even though automated tools for programming practice have been around since the 1960s, such tools have predominantly focused on identifying mistakes rather than suggesting next steps or guidance on fixing problems~\cite{keuning.2019}. Moreover, effective feedback must consider the timing of an intervention, the context, and the learner. Individuals at different competency levels require different types of support, ranging from basic error identification to strategic problem-solving guidance \cite{narciss2008feedback,shute2008formative}.

The emergence of Generative AI (GenAI) tools based on Large Language Models (LLMs) offers promising possibilities, specifically for the automatic generation of individual feedback in the context of programming education \cite{becker.2023a,becker2023generative}. 
However, early studies showed that GenAI tools often provided immediate code solutions to tasks and student questions~\cite{kiesler.2023,kiesler.2023b} instead of providing the scaffolding approach that students prefer for independent learning \cite{denny.2024}. 
Recent evaluations of GenAI feedback have focused on quality metrics such as completeness and accuracy, typically assessed through expert evaluation \cite{kiesler.2023,azaiz.2024, hellas.2023, jacobs.2024} or comparative analysis using additional models \cite{koutcheme.2024a}. \citet{scholl2024analyzing} started to investigate students' use patterns in the context of an introductory course~\cite{scholl2024analyzing,scholl2024noviceprogrammersuseexperience}. However, there is a gap in empirical research on how exactly students engage with GenAI feedback in introductory programming education \cite{stone_exploring_2024}.

To address this gap, we adopt a mixed-methods approach combining eye-tracking, and a think-aloud protocol including semi-structured interviews with \textbf{the goal} of investigating students' attention to GenAI vs. compiler feedback, and to what extent it was helpful to them. We are also looking at different students (inexperienced vs. experienced learners). 
Thereby, this study \textbf{contributes} to an increased understanding of students' perception and use of GenAI feedback in the context of introductory programming.

\section{Related Work}
Timely and personalized feedback is crucial for programming education \cite{jeuring.2022}. Yet, it remains challenging to implement such feedback at scale. While traditional automated feedback systems often focus on simple test case results or compiler errors, GenAI has emerged as a promising approach for generating natural language feedback.

In the past years, a variety of customized GenAI tools have been developed. These include tools for feedback and hint generation like CodeAid~\cite{Kazemitabaar2024CodeAid}, Codehelp ~\cite{liffiton2023codehelp}, SCRIPT~\cite{scholl2025script}, LLM Hint Factory~\cite{xiao2024exploring}, StAP-tutor ~\cite{roest.2023} or (context aware) chatbots like CS50 Duck~\cite{liu2024teaching, cs50finetuning2025} and CodeTutor~\cite{lyu2024evaluating}.
For example, \citet{kiesler.2023} analyzed ChatGPT's feedback on introductory programming tasks, demonstrating its ability to provide detailed explanations and suggestions while identifying significant variations between outputs, and also misleading information. 
A comprehensive evaluation of GPT-4 Turbo further improved our understanding of GenAI's feedback capabilities~\cite{azaiz.2024}. The feedback analysis of feedback to 55 student submissions showed substantial improvements in structure and consistency~\cite{azaiz.2024} compared to earlier models. However, only 52\% of the feedback was fully correct and complete, emphasizing the continuing challenges in generating reliable feedback, particularly for novices~\cite{azaiz.2024}.

Researchers developed more complex approaches to enhance feedback quality based on these initial findings. \citet{phung.2024} introduced a dual-model approach, combining GPT-4 as a tutor with GPT-3.5 as a student validator. 
\citet{jacobs.2024a} used Retrieval Augmented Generation to incorporate lecture materials into GenAI feedback for programming tasks. In their study, students strategically chose between quick, general feedback and more detailed, context-aware responses depending on a problem's complexity. 
Addressing the need for more control over the generated feedback, \citet{lohr.2024c} investigated the generation of specific types of feedback based on established feedback taxonomies \cite{keuning.2019}. Through iterative prompt engineering, they showed that GPT-4 could generate the requested feedback types in 63 out of 66 cases. 

While these studies have significantly advanced our understanding of feedback generation with GenAI, a crucial gap remains regarding how students interact with GenAI during problem-solving.
\citet{prather.2024a} conducted a mixed-method study including interviews, eye tracking, and student observations with 21 novice programmers to investigate the benefits and challenges of using GenAI tools in introductory programming. They found that some students could effectively use these tools to accelerate their work, but others developed an illusion of competence~\cite{raj2021professional} and faced new difficulties. The study by \citet{prather.2024a} identified three metacognitive difficulties related to GenAI: (1) interruption from frequent AI suggestions, (2) misleading code suggestions, and (3) false progression awareness. Their findings suggest that GenAI tools may widen the gap between well-prepared and under-prepared students. Other work reported on problematic uses of GenAI, especially by learners without prior knowledge~\cite{kiesler.2024a}. 
Similarly, Nam et al.~\cite{nam2024using} conclude that the benefit from their GenAI system \say{may vary depending on participants’ backgrounds or skills}.
This study further addresses the research gap \cite{stone_exploring_2024} in understanding how students engage with GenAI feedback in introductory programming education.

\section{Feedback-Generation with the Tutor Kai}
To systematically evaluate the use and perceived helpfulness of GenAI feedback in a programming education context, we developed a custom web application, the Tutor Kai \cite{jacobs.2024, jacobs.2024a, jacobs2025unlimitedpracticeopportunitiesautomated}. This app served as the interface for students to engage with programming tasks and receive GenAI (and compiler) feedback throughout this study. 

\subsection{Web Interface}
\begin{figure}[h!]
  \centering
  \includegraphics[width=0.9\linewidth]{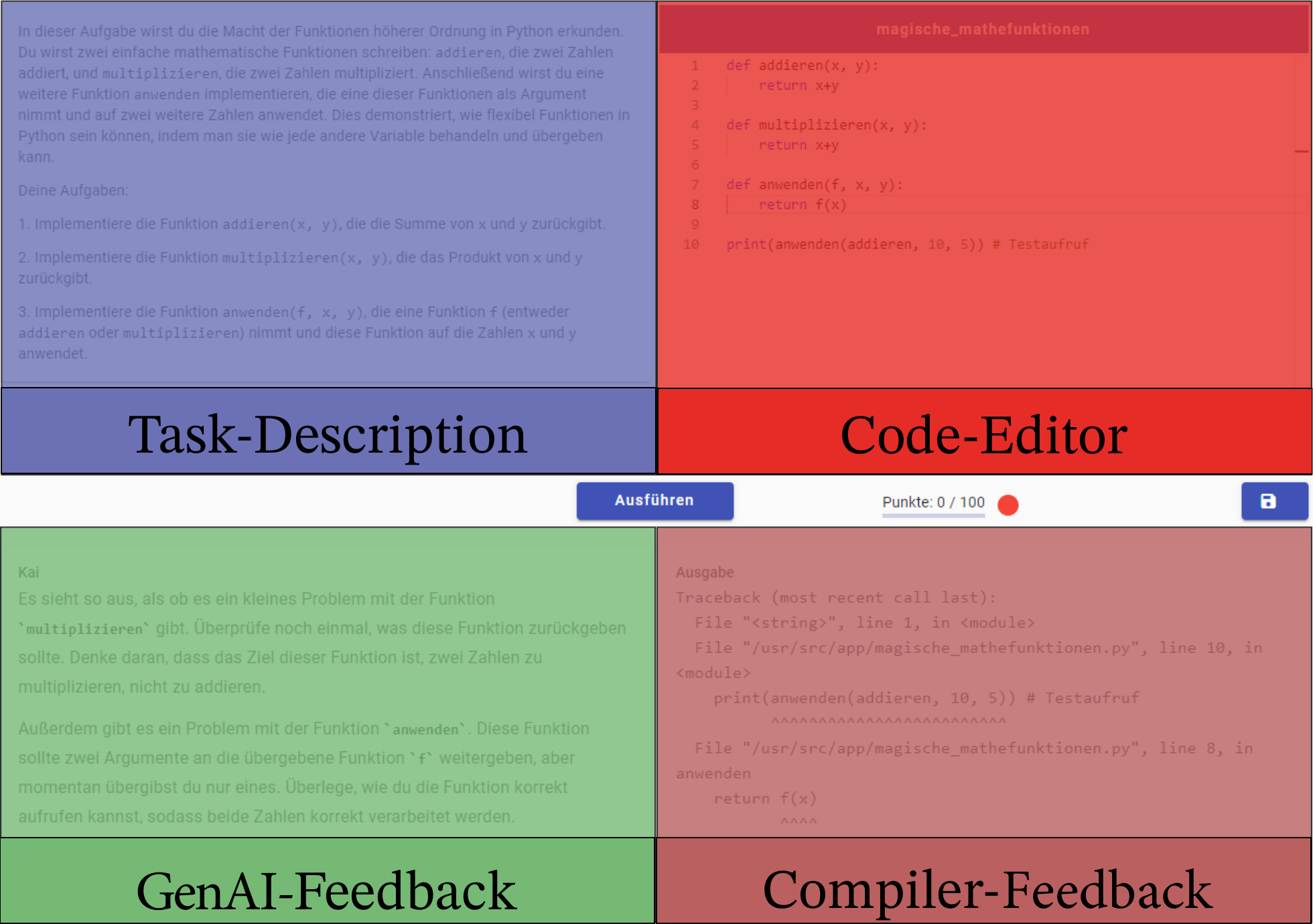}
  \caption{Web application with highlighted AoIs}
  \label{fig_AoIs}
\end{figure}

The web application comprises four Areas of Interest (AoIs), as illustrated in Figure \ref{fig_AoIs}:
\begin{enumerate}[leftmargin=*]
    \item \textit{Task Description} (top left): Provides students with the programming problem to solve.
    \item \textit{Code Editor} (top right): Allows students to write and edit their code.
    \item \textit{GenAI Feedback} (bottom left): Shows generated feedback.
    \item \textit{Compiler Feedback} (bottom right): Displays results of code compilation (compiler output).
\end{enumerate}

\noindent
In the center of Figure \ref{fig_AoIs}, there is a button to execute code. Upon execution, the solution is automatically tested against predefined unit tests for the programming task. Unit-test results are displayed as points. After a code's execution, an additional button appears. It allows students to request feedback. The feedback button remains disabled whenever code changes are made until the code is executed again. This design serves two purposes: (1) to encourage students to make improvements based on compiler feedback and unit test results before requesting GenAI feedback, and (2) the execution results are technically required for generating the feedback. 
Students can make unlimited code executions and GenAI feedback requests.

\subsection{Prompt Engineering for GenAI Feedback}
For generating feedback, we utilized OpenAI's "gpt-4-turbo-2024-04-09" model, with the temperature parameter set to 0 to maximize output consistency. The prompt engineering process was crucial to ensure relevant and pedagogically sound feedback. We developed the prompt iteratively based on our prior work \cite{jacobs.2024}.
The final prompt \cite{supplementarydata.2025} was crafted in Markdown syntax and incorporates comprehensive contextual information, including the task description, the student's current program code, the compiler output, unit test results, and the expected length (6 sentences max.). 
We employed a zero-shot prompting approach combined with role and style prompting techniques \cite{schulhoff.2024}. 

\section{Methodology}
This study is guided by the following research questions (RQs):
\begin{enumerate}
    \item[RQ1]  \textit{How much attention do learners of programming pay to GenAI feedback compared to compiler feedback?}
    \item[RQ2] \textit{To what extent is GenAI feedback helpful for programming learners?} 
\end{enumerate}
\noindent
For both RQs, we particularly investigate differences between students with and without prior programming experience.

A combination of eye-tracking with think-aloud protocols and semi-structured interviews was chosen to capture both, observable behavioral patterns and self-reported perceptions.  
Eye-tracking technology provided objective measures of attention allocation in the web application, while think-aloud protocols offered individual insights into decision-making and problem-solving processes. Interviews allowed for retrospective reflection and clarification. This triangulation of methods supports a comprehensive understanding of students' perceptions of the offered feedback.

The interview guide, programming tasks, think-aloud instructions, prompt used for GenAI feedback, GenAI and compiler feedback examples are available in an online repository~\cite{supplementarydata.2025}.

\subsection{Data Collection}
Eye-tracking technology~\cite{busjahn.2011, busjahn.2014} was used to measure the distribution of visual attention across different Areas of Interest (AoIs). To investigate the helpfulness of GenAI and compiler feedback for inexperienced and experienced learners, we used the think-aloud method \cite{solomon.1994}. This method is common to gain insights into cognitive processes during, e.g., debugging \cite{whalley.2023}, code refactoring \cite{oliveira.2024}, or feedback processing \cite{kiesler.2023c}.

\subsubsection{Participants}
The study was conducted at a university at the beginning of the summer semester of 2024. It involved students enrolled in the ``Object-Oriented and Functional Programming'' (OOFP) course, who are typically first and second-semester computer science students. All course participants (ca.\,200) were invited via email and in-person announcements during tutorial sessions. A compensation of 20€ was offered for participation. 

\subsubsection{Task Design} Six Python programming tasks were specifically developed for this study. The tasks were carefully designed to align with the course curriculum and progress in complexity, covering fundamental programming concepts including variable operations, control structures, loops, and recursion (all tasks are available in our supplementary data repository \cite{supplementarydata.2025}).
To ensure appropriate difficulty levels and time, all tasks were piloted with two student tutors who had previously completed the course. They took about 40 minutes to complete all the tasks.

\subsubsection{Setting}
The study took place in a controlled laboratory environment equipped with a 27-inch monitor and standardized input devices (mouse and keyboard). The primary eye-tracking apparatus consisted of Tobii Pro Glasses 3, complemented by Tobii Pro Lab analysis software. This setup was chosen based on its established reliability in programming education research, with Tobii technology being used in 55\% of related studies \cite{obaidellah.2019}. To ensure data collection continuity, we implemented a backup system using the Tobii Eye Tracker 5. All participants consented after being briefed about data protection, recording procedures, and data processing.

\subsubsection{Procedure}
Each participant was scheduled for a two-hour session, which was structured as follows:

\begin{enumerate}
    \item Introduction (15-30 minutes):
    \begin{itemize}
        \item Think-aloud protocol instruction and practice
        \item Eye-tracking equipment familiarization and calibration
        \item Informed consent process
    \end{itemize}
    \item Task Phase / Problem-Solving (60 minutes):
    \begin{itemize}
        \item Completion of programming tasks
        \item Continuous think-aloud verbalization
        \item Eye-tracking and screen recording
        \item Video and audio recording from front and side angles
    \end{itemize}
    \item Interview Phase (15-30 minutes):
    \begin{itemize}
        \item Clarification of observed behaviors
        \item Reflection of experiences
    \end{itemize}
\end{enumerate}
An instructor remained present throughout the session to address technical issues and provide organizational guidance if needed, while assuring minimal intervention during problem-solving.

\subsection{Data Analysis}
RQ1 is addressed by analyzing the eye-tracking data of each student and mapping it to the Areas of Interest. To answer RQ2, we triangulate the eye-tracking data (and respective AoIs) with students' problem-solving steps to see whether the GenAI feedback, respectively the compiler messages, are indeed helping them succeed with the task. We further analyze instances where GenAI feedback was not helpful. The semi-structured interviews add students' self-reported level of helpfulness for GenAI feedback. 

\subsubsection{Eye-Tracking Analysis}
To analyze students' attention patterns, we defined four AoIs based on the key interface elements of the Tutor Kai: task description, code editor, GenAI feedback, and compiler feedback (see Figure \ref{fig_AoIs}). We mapped gaze data from the eye-tracking glasses' video recordings to these AoIs using assisted mapping features of the Tobii Pro Lab analysis software. For participants S01, S02, and S10, equipment compatibility issues required using a Tobii Eye Tracker 5 as an alternative to the Tobii Pro Glasses 3. For them, we manually mapped their gaze data using frame-by-frame analysis. We analyzed fixation time \cite{obaidellah.2019} for each AoI across all participants, providing a quantitative measure of fixation distribution. To enable further analysis, we exported and synchronized the gaze data recordings with screen captures and external camera footage, creating synchronized videos for a detailed analysis of each participant (i.e., gaze visualization).

\subsubsection{Think-Aloud Protocol Analysis and Triangulation with Eye-Tracking}
First, the audio was transcribed for the detailed analysis of the think-aloud study. Then, the transcript was linked with the synchronized videos for evaluation in MAXQDA~\cite{kuckartz2019analyzing}. All instances of code being executed or feedback requests during the \textit{task phase} were marked. Then, we analyzed students' fixated AoIs immediately before feedback was generated to determine what they were looking at (i.e., compiler feedback, task description, etc.). 

As a next step, students' progress was coded w.r.t. their improvements (or the lack of them) to determine the objective helpfulness of the feedback and answer RQ2. We used the following categories (data sources are indicated in brackets): 

\textbf{Has helped:} Information in the compiler or GenAI feedback was read (gaze visualization). Subsequently, either a thought was verbalized (think-aloud) that explicitly indicated the information helped, or the code was improved in a way that directly related to the provided information. 
E.g., a participant read the GenAI feedback ``[...] It appears your function has not yet implemented a return statement [...]'' and verbalized ``of course [...] return [...]'' while simultaneously correcting the code. Even without the verbalization, this instance would have been coded as \textbf{helpful} because the gaze visualization confirmed the participant had read the relevant GenAI feedback section before adding the return statement.

\textbf{Has not helped:} Information in the compiler or GenAI feedback was read (gaze visualization). Participants did not verbalize anything that indicated the information helped (think-aloud). No code improvements were made related to the provided information, e.g., some students read the compiler feedback but immediately requested GenAI feedback instead of acting on it. 
For future improvements in feedback generation, we also categorized why GenAI feedback was unhelpful.

\textbf{Has not been read:} Information in the compiler or GenAI feedback was not read (gaze visualization). For example, GenAI feedback was requested before reading the compiler feedback, or during GenAI feedback generation. Or the student made improvements independently and immediately requested new GenAI feedback.

\subsubsection{Interview Analysis}
We conducted semi-structured interviews to gain more insights into participants' perceptions regarding the helpfulness of the compiler and GenAI feedback. The complete interview guide is available~\cite{supplementarydata.2025}. 
In particular, we wanted to let students rate (1) task difficulty (10-point-Likert scale, very easy -- very difficult) and (2) the helpfulness of GenAI feedback (10-point-Likert scale, not helpful at all -- very helpful).
Responses are quantitatively analyzed and reported. In addition, we assessed interferences from the think-aloud protocol and eye-tracking equipment if explicitly uttered by participants in response to dedicated questions. 
Finally, we analyzed whether or not students had any prior programming experience resulting in a binary distinction.

\section{Results}

\subsection{Student Sample}
11 students voluntarily participated and completed the study (S01-S11). 
The students recruited for this study were all familiar with the tool, as it was regularly used in the OOFP course. 
7 identified themselves as males, 4 as females. Their mean age was 21 (SD=2.79). The analysis of students' prior programming experience quickly revealed that there were inexperienced and experienced students among the recruits (as expected). 
6 students were \textit{inexperienced}, reporting little or no prior programming experience. The remaining 5 students were classified as \textit{experienced}, as they reported prior programming experience. This included exposure to programming through high school computer science courses or completion of programming-related coursework in previous academic terms.

It should be noted that the students enrolled in the OOFP course were generally diverse in terms of prior knowledge. An initial survey (N=97) within the course revealed that 43 students (about half) had no prior knowledge. Hence, we assume that our sample is representative w.r.t. that diversity.

Our analysis showed distinct patterns in both task completion rates and perceived difficulty levels between these two experience groups. Experienced students completed an average of 4.2 tasks, compared to 1.8 tasks for inexperienced students. Students' subjective assessments of task difficulty, collected through semi-structured interviews using a 10-point Likert scale (1 = very easy, 10 = very difficult), aligned with these performance differences: experienced students rated the tasks as moderately challenging (M = 4.2), while inexperienced participants consistently reported higher difficulty levels (M = 7.0). This data thus confirms students' self-reported prior programming experiences (or the lack of them).

\subsection{RQ1: Students' Attention Distribution} 

The code editor (53.91\%) received the longest fixation time overall, meaning among all 11 students. 
Compared to the other AoIs, GenAI feedback accounted for 23.79\% of the total fixation time, which is more than three times longer than the compiler feedback received in total (7.00\%). Compared to the task description (15.29\%), the fixation time on GenAI feedback was also higher. 

\begin{figure}[h!]
  \centering
  \small
  \includegraphics[width=0.9\linewidth]{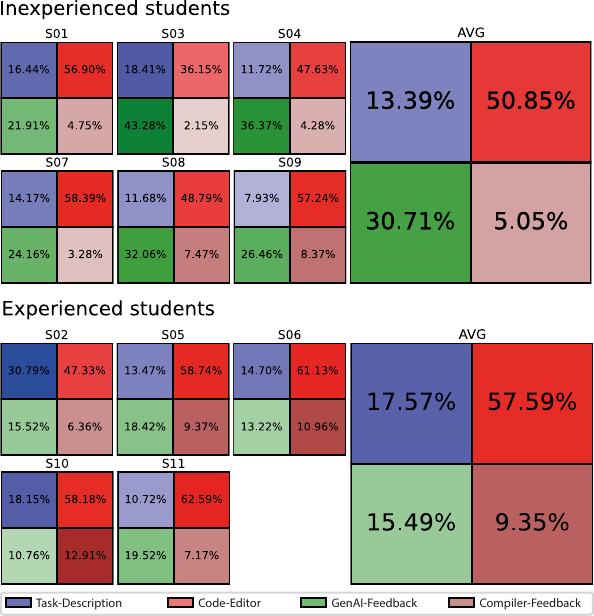}
  \caption{Fixation Times on Areas of Interest}
  \label{fig_EyeTracking}
\end{figure}
\noindent
Comparative analysis reveals that experienced students spent 15.49\% of the time on the GenAI feedback, which is approximately half as much as inexperienced students spent on the GenAI feedback (30.71\%). 
The inverse pattern emerged for compiler feedback, where experienced students exhibited nearly double the fixation time (9.35\%) compared to inexperienced students (5.05\%). 
Regarding the task description and code-editor area, experienced students paid some more attention to them, compared to the less experienced students (see Figure \ref{fig_EyeTracking}). 

\begin{figure*}[htb]
  \centering
  \includegraphics[width=0.95\linewidth]{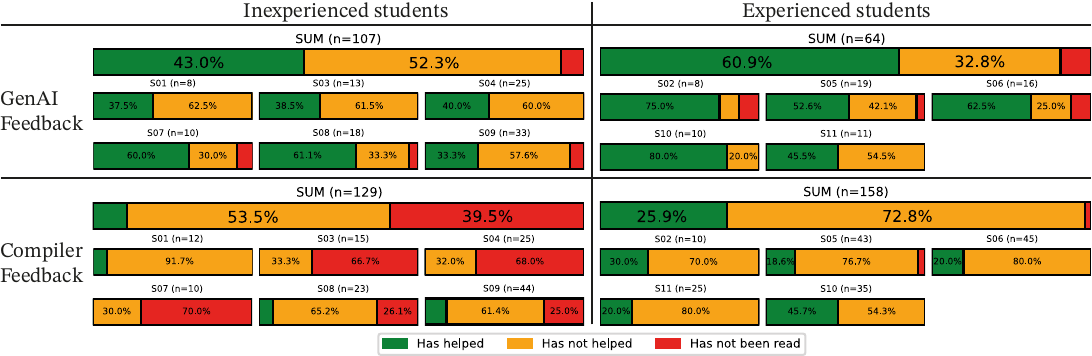}
  \caption{Resulting Helpfulness of GenAI and Compiler Feedback for Experienced and Inexperienced Students (RQ2)}
  \label{fig_Feedhack_Utilization}
\end{figure*}

\subsection{RQ2: Helpfulness of the GenAI Feedback}
First of all, we looked at the fixated AoI, immediately before feedback was generated to determine what students were looking at (see Table \ref{tab:FixiationBeforeFeedback}). The code editor and compiler feedback were each fixated right before about 50\% of the 171 GenAI feedback requests. The task description was fixated only once. Inexperienced students fixated on the compiler feedback less often (36.45\%) compared to experienced students (68.75\%).

\begin{table}[tb]
\small
\centering
\caption{Fixiated AoI before GenAI Feedback Request}
\begin{tabular}{l c c c}
\toprule
 & \shortstack{Task-\\Description} & \shortstack{Code-\\Editor} & \shortstack{Compiler-\\Feedback} \\
\midrule
Inexperienced students & 0 (0\%) & 68 (63.55\%) & 39 (36.45\%)\\
Experienced students & 1 (1.56\%) & 19 (29.69\%) & 44 (68.75\%) \\
All students & 1 (0.58\%) & 87 (50.88\%) & 83 (48.54\%) \\
\bottomrule
\end{tabular}
\label{tab:FixiationBeforeFeedback}
\end{table}

\begin{table}[tb]
  \caption{Results of the Semi-Structured Interviews}
  \label{tab:interviews}
  {\footnotesize
  \begin{tabular}{l@{\hspace{6pt}}c@{\hspace{2pt}}c@{\hspace{2pt}}c@{\hspace{2pt}}c@{\hspace{2pt}}c@{\hspace{2pt}}c@{\hspace{4pt}}c@{\hspace{2pt}}c@{\hspace{2pt}}c@{\hspace{2pt}}c@{\hspace{2pt}}c}
    \toprule
    & \multicolumn{6}{c}{Inexperienced students} & \multicolumn{5}{c}{Experienced students} \\
    \cmidrule(lr){2-7} \cmidrule(lr){8-12}
    & S01 & S03 & S04 & S07 & S08 & S09 & S02 & S05 & S06 & S10 & S11 \\
    \midrule
    \multicolumn{1}{l}{Solved tasks count (of 6)} & 1 & 2 & 2 & 2 & 2 & 2 & 5 & 4 & 5 & 4 & 3\\
    \midrule
    \multicolumn{12}{l}{Likert Scale Items (1-10):} \\
    \multicolumn{1}{l}{Perceived Task Difficulty} & 5 & 8 & 7 & 7 & 8 & 7 & 1 & 6 & 5 & 4 & 4\\
    \multicolumn{1}{l}{Feedback Rating} & 8 & 7.5 & 10 & 10 & 10 & 8 & 9 & 8 & 7 & 7.5 & 10\\
    \midrule
    \multicolumn{1}{l}{Prior programming experience} & \Circle & \Circle & \Circle & \Circle & \Circle & \Circle & \CIRCLE & \CIRCLE & \CIRCLE & \CIRCLE & \CIRCLE\\ \midrule
    \multicolumn{12}{l}{Feedback:} \\
    ...was easy to understand & \CIRCLE & \CIRCLE & \CIRCLE & \CIRCLE & \CIRCLE & \Circle & \CIRCLE & \CIRCLE & \CIRCLE & \CIRCLE & \CIRCLE \\
    ...was sometimes too vague & \CIRCLE & \CIRCLE & \CIRCLE & \CIRCLE & \CIRCLE & \CIRCLE & \Circle & \CIRCLE & \CIRCLE & \CIRCLE & \Circle \\
    ...had too complex vocabulary & \Circle & \Circle & \CIRCLE & \CIRCLE & \CIRCLE & \Circle & \Circle & \CIRCLE & \Circle & \Circle & \CIRCLE \\
    ...increased motivation & \CIRCLE & \CIRCLE & \CIRCLE & \CIRCLE & \CIRCLE & \CIRCLE & \CIRCLE & \CIRCLE & \CIRCLE & \CIRCLE & \CIRCLE \\
    \bottomrule
  \end{tabular}
   }
\end{table}

There were 171 GenAI feedback outputs, 107 for inexperienced and 64 for experienced students. 49.7\% of them helped advance problem-solving processes to solve the tasks correctly (see Figure \ref{fig_Feedhack_Utilization}). The information in the 287 compiler feedback messages was substantially less helpful (overall in just 17.4\%). 

Our analysis reveals substantial differences between the two groups of students. 
As shown in Figure \ref{fig_Feedhack_Utilization}, GenAI feedback was more helpful for experienced students (60.9\%) compared to inexperienced students (43.0\%). Experienced students requested 28\% less GenAI feedback on average (only 64 in sum) compared to 107 GenAI feedback requests from inexperienced students.
The analysis of the synchronized think-aloud protocols showed that the compiler feedback was almost four times more helpful for experienced students (25.9\%) than for inexperienced students (7\%). The eye-tracking data confirms that 39.5\% of the 129 compiler feedback messages were not read by inexperienced students (lower left quadrant of Figure \ref{fig_Feedhack_Utilization}). The fixation times on AoIs further confirm that inexperienced students fixated only half as long on the compiler feedback compared to experienced students (see Figure \ref{fig_EyeTracking}).
Instead of reading the compiler feedback first, inexperienced students often requested GenAI feedback immediately, explaining the higher number of GenAI feedback requests. 

Our analysis of the 77 instances where GenAI feedback was classified as \say{has not helped} revealed four categories of issues: (1) 25 cases contained a range of various causes that opposed further distinct categorization, such as student misinterpretation of the feedback. (2) In 23 instances (22 for inexperienced, 1 for experienced students), the feedback was incomprehensible to the student. This occurred when the GenAI feedback referred to fundamental conceptual knowledge (e.g., 'loop') unfamiliar to the student without sufficient explanation or illustrative examples. (3) In 17 cases (12 for inexperienced, 5 for experienced students), students grasped the semantic solution but were unfamiliar with the necessary Python syntax, which the GenAI feedback failed to provide (e.g., through a code example). 
They expressed that it was not the feedback they needed. (4) In 13 instances, the error localization in the GenAI feedback lacked sufficient precision (e.g., indentation errors were indicated but remained difficult for students to identify).

Students' overall perception of the GenAI feedback was positive. In the interviews, students rated its helpfulness at an average of 8.59 on a 10-point Likert scale (1 = not helpful at all, 10 = very helpful, see Table \ref{tab:interviews}). This high rating was consistent across both experience groups, with no participant rating the feedback below 7 (see Table \ref{tab:interviews}). Yet, the in-depth analysis of think-aloud protocols and triangulation with the eye-tracking data suggests a more complex relationship between perceived and actual GenAI feedback helpfulness.
The remaining interview results regarding students' retrospection were somewhat inconclusive (see Table \ref{tab:interviews}). Ten out of eleven students described the GenAI feedback as generally understandable, but also sometimes not precise enough. Five students mentioned that the GenAI feedback sometimes used too complex technical vocabulary. All students stated that the GenAI feedback increased their motivation to solve programming tasks.

\section{Discussion}
Our findings on GenAI feedback reflect broader patterns emerging in computing education due to the advent of GenAI tools. The gap observed between experienced and inexperienced students in utilizing GenAI feedback has significant implications. Recent work already reported how the introduction of GenAI tools coincided with dramatically increased failure rates in introductory programming courses~\cite{kiesler.2024a}. This divergence in outcomes aligns with the observation that while some students can effectively use GenAI to ``accelerate'' their programming work, others may develop an ``illusion of competence''.  AI tools seem to introduce additional metacognitive difficulties, and students are often unaware of them~\cite{prather.2024a}.

Furthermore, the results indicate that inexperienced students often do not read the compiler feedback. Instead, they read the GenAI feedback. This is not surprising, since programming error messages are generally problematic for students - even though they play an essential role as feedback agents~\cite{becker.2023a}. If these inexperienced students perpetuate this strategy, they may not learn to interpret compiler outputs correctly. At the same time, it could be argued that with the ubiquity of GenAI support, interpreting compiler outputs may no longer be necessary in the future. After all, compiler messages can be directly explained by an LLM \cite{leinonen.2023, santos_not_2024}.
While \citet{reeves.2024} suggest natural language programming through GenAI as next logical step in programming abstraction, our results indicate that students still need to develop basic programming knowledge and skills to comprehend GenAI feedback. 

Successful integration of GenAI into programming education may require a balanced approach combining traditional foundational learning with gradual exposure to GenAI-assisted programming techniques. In this context, \citet{keuning.2024a} suggest a mastery learning approach, which allows students to develop foundational skills while gradually incorporating GenAI tools. 
Ultimately, we need to consider which competencies students should develop. 

\section{Threats to Validity}
This study's methodological limitations should be noted. 
For example, the think-aloud protocol and eye-tracking technology may have disrupted the problem-solving process. In the interviews, three students reported the think-aloud protocol as disruptive, noting difficulties in verbalizing their thoughts or finding the selection of which thoughts to verbalize burdensomely. 
The observation setting itself affected some participants' comfort levels. Four students reported feeling embarrassed about their code and thoughts while being observed. 
Regarding the eye-tracking equipment, two students experienced physical distractions: one reported the glasses became warm during use, another one reported distractions caused by the glasses' sensors. 
A technical limitation is due to the non-deterministic LLM for GenAI feedback generation; the output generation of LLMs remains intransparent~\cite{schulhoff.2024}. 
The nature of the study is qualitative and explorative. With the sample size of 11 students and 171 GenAI feedbacks, we have reached a large enough sample~\cite{boddy2016sample}.

\section{Conclusion}
In this study, we evaluated how 11 students utilize feedback generated by GenAI while solving programming tasks in the Tutor Kai. Our mixed-methods approach, combining eye-tracking, think-aloud protocols, and interviews, revealed the following results:

\begin{enumerate}[leftmargin=*]
\item Students dedicated a substantial portion of their visual attention (23.79\%) to the GenAI feedback.
However, there were notable differences between experience levels. Inexperienced students spent nearly twice as much time fixating on the GenAI feedback compared to more experienced peers.
\item Experienced students demonstrated a more effective utilization of the GenAI feedback. They requested it less frequently and derived more benefit from it, with 60.9\% of the GenAI feedback helping to advance their problem-solving, compared to only 43.0\% for inexperienced students.
One reason for this was that inexperienced students did not comprehend 20.6\% (22 out of 107 cases) of the GenAI feedback they received, compared to only 1.6\% (1 out of 64 cases) for experienced students.
\item Inexperienced students often bypassed the compiler feedback entirely, relying solely on the GenAI feedback. This raises concerns about the potential impact on the development of traditional debugging skills.
\end{enumerate}
\noindent
Our findings highlight both the potential of GenAI feedback in programming education and the need for tailored approaches based on students' experience. Prior research~\cite{denny.2024} suggests that this is also important to students. 
Future research should focus on developing feedback systems that provide feedback adapted to students' prior knowledge and experience.
Moreover, they should be capable of adapting to students' individual informational needs. As this work has shown, experienced and inexperienced students seem to engage differently with compiler vs. GenAI feedback. Hence, this study should be repeated with a greater number of participants. More importantly, we need to develop instructional methods for diverse student groups with and without prior programming knowledge so they learn how to engage with both, compiler and GenAI feedback.

\bibliographystyle{ACM-Reference-Format}
\bibliography{Student_Engagement_with_GenAI_Feedback}

\end{document}